\documentclass[sigconf]{acmart}

\AtBeginDocument{%
  }

\usepackage{enumitem}
\usepackage{multirow}
\usepackage{bm}
\usepackage{subfigure}
\usepackage{textcomp}%

\begin{document}

\title{The 2nd Workshop on Human-Centered Recommender Systems}

\author{Kaike Zhang}
\affiliation{%
  \institution{University of Chinese Academy of Sciences}
  \city{Beijing}
  \country{China}
}

\email{kaikezhang99@gmail.com}

\author{Jiakai Tang}
\affiliation{%
  \institution{Renmin University of China}
  \city{Beijing}
  \country{China}
}
\email{tangjiakai5704@ruc.edu.cn}

\author{Du Su}
\affiliation{%
  \institution{Institute of Computing Technology, CAS}
  \city{Beijing}
  \country{China}
}
\email{sudu@ict.ac.cn}

\author{Shuchang Liu}
\affiliation{%
  \institution{Kuaishou}
  \city{Beijing}
  \country{China}
}
\email{liushuchang@kuaishou.com}

\author{Julian McAuley}
\affiliation{%
  \institution{University of California, San Diego}
  \city{La Jolla}
  \country{USA}
}
\email{jmcauley@eng.ucsd.edu}

\author{Lina Yao}
\affiliation{%
  \institution{CSIRO Data61}
  \city{Sydney}
  \country{Australia}
}
\email{lina.yao@data61.csiro.au}

\author{Qi Cao}
\affiliation{%
  \institution{Institute of Computing Technology, CAS}
  \city{Beijing}
  \country{China}
}
\email{caoqi@ict.ac.cn}

\author{Yue Feng}
\affiliation{%
  \institution{University of Birmingham}
  \city{Birmingham}
  \country{UK}
}
\email{y.feng.6@bham.ac.uk}

\author{Fei Sun}
\affiliation{%
  \institution{University of Chinese Academy of Sciences}
  \city{Beijing}
  \country{China}
}
\email{ofey.sunfei@gmail.com}

\renewcommand{\shortauthors}{Kaike Zhang et al.}

\begin{abstract}

Recommender systems shape how people discover information, form opinions, and connect with society. 
Yet, as their influence grows, traditional metrics, e.g., accuracy, clicks, and engagement, no longer capture what truly matters to humans. 
The workshop on Human-Centered Recommender Systems (HCRS) calls for a paradigm shift from optimizing engagement toward designing systems that truly understand, involve, and benefit people. 
It brings together researchers in recommender systems, human-computer interaction, AI safety, and social computing to explore how human values, e.g., trust, safety, fairness, transparency, and well-being, can be integrated into recommendation processes. 
Centered around three thematic axes---Human Understanding, Human Involvement, and Human Impact---HCRS features keynotes, panels, and papers covering topics from LLM-based interactive recommenders to societal welfare optimization. By fostering interdisciplinary collaboration, HCRS aims to shape the next decade of responsible and human-aligned recommendation research.



\end{abstract}



\maketitle

\section{Introduction}

Recommender systems have become essential for managing the exponential growth of online information, profoundly shaping how people make decisions across domains such as e-commerce, media, and social networking. However, their increasing influence also introduces challenges, including filtering bubbles, privacy, and fairness, which impact users' autonomy, trust, and well-being.

To address these issues, recent research has called for the development of Human-Centered Recommender Systems (HCRS)~\cite{konstan2021human, silva2024leveraging}. HCRS emphasizes placing human needs, values, and capabilities at the core of system design, training, and evaluation. While Trustworthy Recommender Systems focus on transparency, robustness, and privacy~\cite{ge2022survey, wang2024trustworthy}, and Responsible Recommender Systems emphasize fairness, accountability, and bias mitigation~\cite{kazienko2024toward}, HCRS adopts a broader human-first perspective that integrates these aspects while going further to design systems that genuinely understand, involve, and benefit humans.
Specifically, HCRS can be organized around three main thematic axes:
\begin{itemize}[leftmargin=*]
    \item \textbf{Human Understanding}: Modeling user intents, cognition, and affect beyond surface interactions, moving beyond clicks toward multidimensional measures such as satisfaction and trust.
    \item \textbf{Human Involvement}: Enabling interactive, co-adaptive, and controllable systems that empower users to guide model updates, express preferences, and co-create content.
    \item \textbf{Human Impact}: Addressing societal and ethical dimensions, including fairness, privacy, robustness, and transparency, and mitigating issues such as echo chambers and filter bubbles.
\end{itemize}

Together, these axes call for a paradigm shift---from optimizing engagement to creating systems that genuinely understand, empower, and respect humans. Bridging recommender systems, HCI, AI safety, and social computing, HCRS aims to advance a new generation of human-aligned and socially responsible recommenders.

\section{Scope and Topics}
\label{sec:scope}

This workshop offers researchers an interdisciplinary platform to present the latest advancements in the rapidly evolving field of Human-Centered Recommender Systems (HCRS). By fostering dialogue between recommender systems, HCI, AI safety, and social computing, HCRS aims to advance the understanding and development of systems that are not only technically effective but also ethically sound, user-centered, and socially responsible. We welcome original submissions that contribute to one or more of the following thematic axes:

\begin{itemize}[leftmargin=*]
    \item \textbf{Human Understanding---Systems that understand humans.}  
    Modeling users' intents, cognition, and affect beyond clicks and ratings. Topics include:
    \begin{itemize}
        \item Intent- and context-aware recommendation, cognitive and affective modeling.
        \item LLM-based user modeling, and human behavior uncertainty.
        \item Beyond-click metrics such as satisfaction, trust, and emotion.
    \end{itemize}

    \item \textbf{Human Involvement---Humans in the loop.}  
    Enhancing user participation, control, and co-adaptation in recommender systems. Topics include:
    \begin{itemize}
        \item Interactive and conversational recommendation, multi-turn and mixed-initiative interaction.
        \item User feedback elicitation, and controllable personalization.
        \item Human–AI co-creation in content-generation scenarios.
        \item User simulation for scalable and safe training and evaluation.
    \end{itemize}

    \item \textbf{Human Impact---Systems that affect humans and society.}  
    Exploring fairness, privacy, transparency, and societal well-being in recommendation. Topics include:
    \begin{itemize}
        \item Fairness and bias mitigation, diversity enhancement, and echo chamber alleviation.
        \item Privacy-preserving recommendation.
        \item Robustness and security.
        \item Transparency and accountability.
        \item Evaluation, auditing, and governance frameworks for responsible recommendation.
        \item Societal-welfare and well-being-oriented optimization.
    \end{itemize}

    \item \textbf{Emerging Cross-Domain Topics.}  
    We also encourage submissions on emerging research directions that bridge disciplines, such as:
    \begin{itemize}
        \item LLM-powered recommender and human preference alignment.
        \item Multi-objective optimization for social good, balancing privacy, diversity, and satisfaction.
        \item Human-centered evaluation methodologies and user studies.
        \item Recommender systems for education, healthcare, etc.
    \end{itemize}
\end{itemize}

\section{Rationale}
The rapid advancement of human-centered recommender systems has catalyzed a wave of innovative research efforts in recent months~\cite{zhang2023robust, wu2024accelerating, wang2024trustworthy, deldjoo2024fairness, yoo2024ensuring, balloccu2024explainable, zhou2024bee}. Consequently, hosting this workshop is essential for fostering discussions on these evolving directions in the field.

\subsection{Relevance}

This workshop on Human-Centered Recommender Systems (HCRS) closely aligns with the vision of The Web Conference to advance technologies that empower people and improve their interaction with the web. 
Recommender systems play a central role in how users explore, access, and engage with online information, making them one of the most visible and influential components of the modern web.
By focusing on human-centered principles, HCRS aims to make recommender systems more trustworthy, transparent, and aligned with human values.
In doing so, the workshop highlights a timely and essential research direction at the intersection of web intelligence, human–AI interaction, and social computing---an area of growing importance for the WWW community.

\subsection{Prior Workshops}
We organized the inaugural Workshop on Human-Centered Recommender Systems (HCRS) at The Web Conference 2025 (WWW~'25) in Sydney, Australia~\cite{zhang20241st}. The event attracted substantial interest, with approximately 40 in-person participants and over 60 remote attendees via Zoom. The first edition received 14 submissions, of which 6 were accepted. Further details are available at \url{https://human-centeredrec.github.io/}.

\subsection{Objectives and Expected Outcomes}
This workshop encourages researchers to propose new theoretical frameworks, interdisciplinary approaches, and perspectives for human-centered recommender systems. We will also advocate for the adoption of advanced technologies, such as large language models, to enhance the human-centric qualities of existing recommender systems. Additionally, participants will be encouraged to develop innovative evaluation frameworks and metrics tailored to assessing these qualities. In the long term, this research focus is expected to drive the evolution of recommender systems into broader domains, fostering their applicability across various contexts and contributing positively to the well-being of society.

\subsection{Target Audience}
The appeal of this workshop lies in its commitment to the continuously evolving landscape of recommender systems. It aims to attract a diverse audience, including researchers, industry experts, and academics. The workshop will provide these stakeholders with a unique forum to share innovative ideas, methodologies, and achievements, fostering interdisciplinary collaboration and exploring new applications. By bringing together a broad spectrum of participants, we hope to stimulate rich discussions and catalyze the advancement of human-centered approaches in the realm of recommender systems.

\subsection{Diversity and Inclusion}

\textbf{Diversity among Organizers}. Our organizing team is characterized by diversity in terms of gender, affiliations, nationality, professional backgrounds, and levels of seniority.
The team comprises members from both academia and industry, representing four major regions---Asia, Oceania, North America, and Europe. 
Notably, women account for one-third of the organizers.
This broad and inclusive composition ensures that the workshop benefits from a rich variety of perspectives and experiences, fostering balanced discussions and globally relevant outcomes.


\textbf{Diversity among Speakers}. We will invite researchers from diverse backgrounds across both academia and industry to deliver keynote talks, ensuring a wide range of perspectives and expertise. We have confirmed keynotes from \textbf{Prof. Zhaochun Ren (Leiden University)} and \textbf{Prof. Asia J. Biega (Max Planck Institute for Security and Privacy)}, and are currently confirming \textbf{Dr. Ed Chi, VP of Research at Google DeepMind}. 

Furthermore, we encourage submissions from scholars and practitioners across the globe, aiming to highlight the work of diverse voices accepted into this workshop.

To promote diversity among our invited panel speakers, we will curate a mix of participants, including academics and industry professionals, representing various levels of seniority and experience in their respective fields. This approach will foster rich dialogue, allowing for a comprehensive exchange of ideas and perspectives that reflect the multifaceted nature of HCRS.

\section{Workshop Program Details}

\textbf{Workshop Program Format.} This workshop will be conducted over a \textbf{half-day} session. We will invite three researchers to deliver 30-minute keynote talks, providing valuable insights into the latest advancements and future prospects of HCRS. 
Besides, we will host a panel discussion featuring several senior researchers, focusing on future directions and challenges in this domain. 

We expect to accept 10--15 papers at this workshop. We will also incorporate a paper presentation session to allow participants to share their research findings and foster academic discourse. The preliminary program schedule is presented in Table~\ref{tab:schedule}.

\begin{table}[t]
  \centering
    \caption{Program Schedule}
    \resizebox{0.4\textwidth}{!}{
\begin{tabular}{lc}
    \toprule
    \textbf{Event} & \textbf{Time} \\
    \midrule
     Opening Remarks from Co-Chairs & 09:00–09:10 \\
     Keynote Talk \#1 followed by Q\&A & 09:10–09:40 \\
     Paper Session \#1 & 09:40–10:00 \\
     Keynote Talk \#2 followed by Q\&A & 10:00–10:30 \\ 
     Tea Break & 10:30–11:00 \\
     Keynote Talk \#3 followed by Q\&A & 11:00–11:30 \\
     Paper Session \#2 & 11:30–12:10 \\
     Panel Discussion \& Closing Remarks & 12:10–12:30 \\
    \bottomrule
    \end{tabular}
    }
  \label{tab:schedule}%
\end{table}

This structured schedule is designed to facilitate dynamic interactions among participants, encouraging networking opportunities and collaborative discussions. We anticipate that the combination of keynote addresses, research presentations, and a panel discussion will provide a comprehensive exploration of human-centered recommender systems, fostering an environment conducive to innovation and the exchange of ideas.

\textbf{Invited Speakers.} We are honored to feature three distinguished keynote speakers who represent complementary perspectives in the fields of human-centered recommender systems:
\begin{itemize}[leftmargin=*]
\item \textbf{Zhaochun Ren (Confirmed)} (Leiden University, Netherlands): Dr. Ren’s research bridges information retrieval, NLP, and recommender systems, focusing on conversational information-seeking and intelligent interaction. His work advances user-centered and explainable recommendation, supported by strong ties between academia and industry.
\item \textbf{Asia J. Biega (Confirmed)} (Max Planck Institute for Security and Privacy, Germany): Dr. Biega leads the Responsible Computing group at MPI-SP, studying fairness, accountability, and digital well-being in data-driven systems. As ACM FAccT 2025 General Co-Chair, she brings vital perspectives on aligning recommendation technologies with human values and social good.
\end{itemize}

\textbf{Program Committee.} We have contacted the following individuals to serve on our Program Committee: Yuan Zhang (ByteDance), Xinyu Lin (National University of Singapore), Chen Xu (Renmin University of China), Yuanhao Liu (Institute of Computing Technology), Xiao Lin (Kuaishou), Jiakai Tang (Renmin University of China), Huizhong Guo (Zhejiang University), Yuyue Zhao (University of Science and Technology of China), Zhiyu He (Tsinghua University) and Zhaolin Gao (Cornel University). We also plan to invite the following individuals, whom we have not yet contacted: Allegra De Filippo (University of Bologna), Fabrizio Silvestri (Sapienza University of Rome), Michael D. Ekstrand (Drexel University), Anshuman Chhabra (University of South Florida), Ivan Srba (Kempelen Institute of Intelligent Technologies).

\section{Call for Papers}
This workshop aims to encourage innovative research on human-centered recommender systems, particularly in the areas of robustness, privacy, transparency, fairness, diversity, accountability, ethical considerations, user-friendly design, and evaluation methods. We summarize the detailed objectives and scope in Section~\ref{sec:scope}.

All submitted papers must be formatted as a single PDF document according to the ACM WWW 2026 template. Manuscripts may range from 4 to 8 pages in length, with unlimited pages allowed for references. Authors are encouraged to determine the appropriate length for their submissions, as there is no distinction made between long and short papers.

All submissions will adhere to a single-blind review policy. Each paper will undergo a rigorous review procedure, with expert peer reviewers assessing submissions based on their relevance to the workshop, scientific novelty, and technical quality.

The important dates for the submission process are as follows:
\begin{itemize}
    \item \textbf{Submission Deadline}: December 18, 2025
    \item \textbf{Paper Acceptance Notification}: January 13, 2026
    \item \textbf{Camera-Ready Submission}: February 2, 2026
\end{itemize}

\section{Organizers}
\begin{itemize}[leftmargin=*]
    \item \textbf{Kaike Zhang} (University of Chinese Academy of Sciences) is a Ph.D. candidate at the University of Chinese Academy of Sciences, under the supervision of Prof. Xueqi Cheng and Prof. Xinran Liu. His research focuses on trustworthy recommender systems. He has published in top-tier conferences such as NeurIPS, SIGIR, SIGKDD, WWW, and RecSys. He has also served as a reviewer and Program Committee member for high-level conferences and journals, including SIGIR, ICLR, WWW, and TKDD. Additionally, \textit{He served as the lead organizer for HCRS@TheWebConf 2025.}
    
    \item \textbf{Jiakai Tang} (Renmin University of China) is currently a Ph.D. student at Renmin University of China, under the supervision of Prof. Xu Chen. His research interests focus on LLM-based recommender systems. He has published nearly 20 papers in top-tier conferences and journals such as KDD, WWW, ACL, and TOIS. He also received the runner-up for the Best Resource Paper Award at CIKM '22. He also serves as a reviewer for data mining-related conferences such as KDD and WSDM.
    
    \item \textbf{Du Su} (Institute of Computing Technology, CAS) is an Assistant Professor at the CAS Key Laboratory of AI Safety, Institute of Computing Technology, Chinese Academy of Sciences. His research interests are in AI safety and security, with a specific focus on technologies for monitoring and assessing AI risks. His goal is to develop an agent-based sandbox environment for algorithm simulation, aimed at assessing algorithms' safety properties from the user's perspective, including privacy, fairness, and filter bubbles. He has published his work in top conferences such as KDD, WWW, and EMNLP. \textit{He has also contributed to the community by co-organizing HCRS@TheWebConf 2025.}
    
    \item \textbf{Shuchang Liu} (Kuaishou) is an Applied Scientist at Kuaishou Technology, Beijing. He earned his Ph.D. in Computer Science from Rutgers University in 2022, under the supervision of Prof. Yongfeng Zhang. His research interests are in recommender systems, reinforcement learning, generative methods, and AI4Science. He has served as a Program Committee member for conferences and journals including AAAI, IJCAI, KDD, RecSys, TOIS, TORS, NeurIPS, and WWW. \textit{He also has contributed to the community by co-organizing HCRS@TheWebConf 2025.}

    \item \textbf{Julian McAuley} (University of California, San Diego) is a Professor at the University of California, San Diego (UCSD). He is a leading researcher in machine learning and recommender systems, has received several prestigious awards, including the SIGIR Test of Time Award, WWW Best Paper Award, and ICWSM Best Paper Award, recognizing his impactful contributions to information retrieval, content moderation, and fashion trend prediction. He has also been honored with research awards from Google, Amazon, and the NSF CAREER Award, highlighting his innovation in improving recommendation algorithms. \textit{He has also made significant contributions to the community by co-organizing workshops, such as WWW 2025 and KDD 2025.}

    \item \textbf{Lina Yao} (CSIRO Data61) is a Senior Principal Research Scientist and Science Lead at CSIRO's Data61, with adjunct and honorary professorships at UNSW, Macquarie University, and UTS. A Senior Member of ACM and IEEE, she serves as Associate Editor for top journals including ACM TOSN, TORS, TALLIP, and IEEE TAI. Her research focuses on generalizable and explainable data mining, with broad applications in recommender systems. \textit{She also served as Workshop Co-Chair for CIKM 2023.}
    
    \item \textbf{Qi Cao} (Institute of Computing Technology, CAS) is an Associate Professor at the ICT, CAS. Her research interests include social media analysis, robustness of recommender systems, and fairness auditing in machine learning. She has published over 30 papers in high-level conferences and journals such as KDD, WWW, NeurIPS, SIGIR, and WSDM. She was nominated for the Chinese Information Processing Society of China (CIPS) Outstanding Doctoral Dissertation Award. She has also served as a Program Committee member for conferences including AAAI, WWW, NeurIPS, and ICLR, and as a reviewer for journals such as TKDE, TOIS, and TKDD. \textit{She has also contributed to the community by co-organizing HCRS@TheWebConf 2025.}

    \item \textbf{Yue Feng} (University of Birmingham) is an assistant professor at the School of Computer Science of University of Birmingham. She obtained her Ph.D. at University College London (UCL). Her research interests lie in natural language processing and information retrieval, with a particular emphasis on LLM–powered AI agents. She has published over 40 papers in top-tier conferences, including WWW, SIGIR, ACL, etc. She has won the Amazon Alexa Prize TaskBot Challenge at 2021. She also co-organized \textit{The 1st EReL-MIR Workshop@WWW 2025} and served as \textit{the Proceedings Chair for SIGIR-AP 2025}.

    \item \textbf{Fei Sun} (University of Chinese Academy of Sciences) is an Associate Professor at the University of Chinese Academy of
Sciences. His research spans recommender systems and natural language processing, with a particular emphasis on safety-oriented topics. He has published over 60 papers in leading conferences and journals, including WWW, SIGIR, and TOIS, and was awarded the Best Paper Runner-Up at RecSys 2019 and the Best Paper Award at KnowFM@ACL 2025. Before joining academia, he spent five years at Alibaba focusing on e-commerce recommendation. \textit{He has also contributed to the community by co-organizing workshops, including NLP4REC@WSDM 2020 and HCRS@TheWebConf 2025.}
\end{itemize}

\begin{acks}
This work was supported by the Strategic Priority Research
Program of the Chinese Academy of Sciences
(XDB0680201), the National Natural Science Foundation of China (62472409), and the Beijing Natural Science Foundation (4252023).
\end{acks}

\bibliographystyle{ACM-Reference-Format}
\bibliography{ref}

\end{document}